\documentclass[conference]{IEEEtran}
\usepackage[T1]{fontenc}
\usepackage[latin9]{inputenc}
\usepackage{geometry}
\usepackage{color}
\usepackage{float}
\usepackage{amsmath}
\usepackage{amsthm}
\usepackage{amssymb}
\usepackage{graphicx}
\usepackage{esint}

\makeatletter

\providecommand{\tabularnewline}{\\}
\floatstyle{ruled}
\newfloat{algorithm}{tbp}{loa}
\providecommand{\algorithmname}{Algorithm}
\floatname{algorithm}{\protect\algorithmname}

\theoremstyle{plain}
\newtheorem{thm}{\protect\theoremname}
\theoremstyle{plain}
\newtheorem{lem}[thm]{\protect\lemmaname}

\usepackage{caption}
\usepackage{amsmath}

\usepackage{caption}
\usepackage{cite}

\IEEEoverridecommandlockouts

\@ifundefined{showcaptionsetup}{}{%
 \PassOptionsToPackage{caption=false}{subfig}}
\usepackage{subfig}
\makeatother

\providecommand{\lemmaname}{Lemma}
\providecommand{\theoremname}{Theorem}

\begin{document}

\title{Cache Size Allocation in Backhaul Limited Wireless Networks}

\author{Xi Peng$^{*}$, Jun Zhang$^{*}$, S.H. Song$^{*}$ and Khaled B.
Letaief$^{*\dagger}$, \emph{Fellow}, IEEE\\
{\normalsize{}$^{*}$Department of ECE, The Hong Kong University of
Science and Technology}\\
{\small{}E-mail: $^{*}$\{xpengab, eejzhang, eeshsong, eekhaled\}@ust.hk}\\
{\normalsize{}$^{\dagger}$Hamad bin Khalifa University, Doha, Qatar}\\
{\small{}E-mail: $^{\dagger}$kletaief@hbku.edu.qa}}
\maketitle
\begin{abstract}
Caching popular content at base stations is a powerful supplement
to existing limited backhaul links for accommodating the exponentially
increasing mobile data traffic. Given the limited cache budget, we
investigate the cache size allocation problem in cellular networks
to maximize the user success probability (USP),\textcolor{black}{{}
taking wireless channel }statistics\textcolor{black}{, backhaul capacities
and} file popularity distributions\textcolor{black}{{} into consideration}.
The USP is defined as the probability that one user can successfully
download its requested file either from the local cache or via the
backhaul link. We first consider a single-cell scenario and derive
a closed-form expression for the USP, which helps reveal the impacts
of various parameters, such as the file popularity distribution. More
specifically, for a highly concentrated file popularity distribution,
the required cache size is independent of the total number of files,
while for a less concentrated file popularity distribution, the required
cache size is in linear relation to the total number of files. Furthermore,
we study the multi-cell scenario, and provide a bisection search algorithm
to find the optimal cache size allocation. The optimal cache size
allocation is verified by simulations, and it is shown to play a more
significant role when the file popularity distribution is less concentrated.
\end{abstract}

\footnote{This work is supported by the Hong Kong Research Grant Council under
Grant No. 610113. }

\section{Introduction}

The demand for mobile data is continuously increasing, which results
from the prevalence of smart devices, as well as subscribers' desire
for high-quality and low-latency services, such as high definition
(HD) video on demand. It is predicted that global cellular data traffic
will increase nearly tenfold in the next five years, and around 75
percent of the world\textquoteright s mobile data traffic will be
video by 2019 \cite{Cisco}. Such rapid growth has been driving operators
to provide more and more network capacity, which is achieved by enhancing
spectral efficiency and expanding the spectrum. But it is still far
from enough to catch up with the data traffic demand.

One approach to help meet the demand is cell densification. By deploying
a large number of small cells, the network capacity can be tremendously
increased \cite{ChangSmallCell}. However, with cell densification,
a new challenge for backhaul is raised--supporting the aggregate data
rate of all users with a reliable connectivity from the core network
to the small cell base stations (BSs) \cite{Backhaul,5GBackhaul}.
As a status quo, backhaul has become a bottleneck for cellular systems,
since most of the existing backhaul links are of low capacity and
often cannot satisfy the rate requirements. For example, the current
average DSL download data rate in the USA is around 5 Mbps, while
a user needs a speed of at least 2 Mbps to watch HD videos smoothly.
Hence with a large number of users, the network is easy to congest
ad break down. Since it is labor-intensive and expensive to establish
new backhaul infrastructures and unrealistic to lay high-capacity
fibers for every small cell BS, new solutions are needed to overcome
the backhaul limitations.

With the benefits of alleviating the backhaul burden and avoiding
potential congestion, caching popular content at local BSs for backhaul
limited networks has emerged as a cost-effective solution \cite{CacheSmallCell,cache_coding_Placement}.
Local caching can be very effective if a small subset of requested
files have high popularity. \textcolor{black}{Recently, lots of efforts
have been focused on cooperative transmission strategies with caching
\cite{MyPIMRC,cacheLiuAn}, caching placement strategy design \cite{cache_coding_Placement,CachePlacement_bhAware},
and coding design for caching \cite{CodedCaching,DecentralizedCodedCaching}.
However, the caching deployment, more specifically, the cache size
allocation, has received less attention. Although the cost of general-purpose
storage devices keeps decreasing, given the actual budget, the caching
capacity deployed at each BS will not be arbitrarily large. The cost
efficiency of caching should be carefully investigated, and the cache
size allocation for BSs within the network should be optimized. Moreover,
the interplay between the cache storage capacity and backhaul capacity
should be studied.}

The issue of cache size allocation has been well discussed in wired
networks. Such allocation is conducted among network nodes or routers
with regard to various performance metrics, such as the probability
of successful recovery \cite{StorgeAlloc_Reliability} and the remaining
traffic flow in the network \cite{wiredCacheAlloct}. It has been
observed that the network topology and content popularity are two
important factors affecting the cache capacity allocation, but for
wireless caching networks, more complicated factors need to be considered.
Gu \emph{et al}. \cite{storageallocation_transcost} discussed the
storage allocation problem with a given backhaul transmission cost.
But this work assumed error-free transmission and neglected the physical
layer features. In practical cellular systems, BSs will see different
wireless channel conditions and have different types of backhaul links.
Therefore, effective cache size allocation considering these factors
is necessary and crucial for caching deployment.

In this paper, we study the cache size allocation problem in backhaul
limited wireless networks. In particular, we adopt the user success
probability (USP) as the performance metric, on which we demonstrate
the impacts of the wireless channel statistics, backhaul conditions
and file popularity distributions. We first derive the closed-form
expression of the USP in the single-cell scenario and show the backhaul-cache
tradeoff. Also, we reveal the influence of the file popularity distribution
on the required cache size. Then we optimize the cache size allocation
in the multi-cell scenario. Simulation results show that the proposed
allocation algorithm outperforms the uniform allocation, and the performance
gain is significant, especially for less concentrated file popularity
distributions.

\section{System model}

\begin{figure}
\begin{centering}
\subfloat[A single-cell scenario. ]{\centering{}\includegraphics[width=0.33\textwidth]{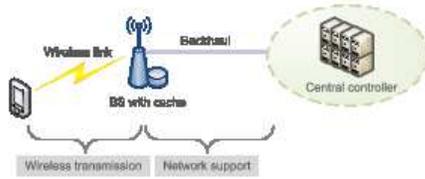}}
\par\end{centering}

\begin{centering}
\subfloat[A multi-cell scenario.]{\centering{}\includegraphics[width=0.33\textwidth]{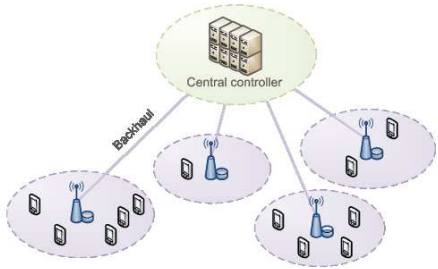}}
\par\end{centering}

\caption{\label{fig:The-architecture}The architecture of a backhaul-limited
caching network. BSs are connected to the central controller through
separate low-capacity backhaul links. Caches can be allocated to BSs
as a supplement to the backhaul.}

\end{figure}

In this paper, we consider a downlink cellular network model consisting
of $N_{b}$ cells, each of which has a base station (BS) and multiple
mobile users, as shown in Fig. \ref{fig:The-architecture}. The radius
of cell $j,\forall j\in\left\{ 1,\dots,N_{b}\right\} $ is denoted
as $R_{j}$, and there is a local BS $j$ located at its center and
$U_{j}$ mobile users uniformly distributed within it. Through separate
backhaul links, BSs are connected to the central controller, which
stores the whole file library. The backhaul capacity of BS $j$ is
denoted as $c_{B,j}$ (bps), and the cache size allocated to BS $j$
is denoted as $c_{S,j}$ (bits).

\subsection{File Request and Caching Model}

The file library has $F$ files, which are assumed to be of the same
length $L$ (bits). Note that the same analysis can be applied to
the case of unequal file sizes, since files can be divided into chunks
of equal length. We do not consider coded caching, and every file
is stored as an entire piece. The popularity of file $f,\forall f\in\left\{ 1,\dots,F\right\} $,
can be characterized by a probability mass function $p\left(f\right)$,
which can be predicted based on collected data \cite{gill2007youtube,Cha07Youtube},
and thus can be regarded as known a priori. Without loss of generality,
the files are sorted in the descending order of popularity, i.e.,
$p\left(1\right)\geq p\left(2\right)\geq\cdots\geq p\left(F\right)$.
Mobile users are assumed to make independent requests according to
$p\left(f\right)$. Since cooperative caching is beyond the scope
of our discussion here, the optimal caching strategy for each BS is
to cache the most popular files, and hence the cache hit ratio of
cell $j$ can be calculated as
\begin{equation}
h_{j}=\sum_{f=1}^{s_{j}}p\left(f\right),\label{eq:hit ratio}
\end{equation}
where $s_{j}\triangleq\left\lfloor \left.c_{S,j}\right/L\right\rfloor $
is defined as the normalized cache size of BS $j$.

\subsection{File Delivery Model}

The delivery of a file depends on two parts: the wireless transmission
and the network support (consisting of backhaul transport and cache
storage), which will be discussed in detail in this subsection.

We assume no interference among users; e.g., they can be served by
different subcarriers with orthogonal frequency-division multiple
access (OFDMA). The wireless downlink rate of a user depends on its
signal-to-noise ratio (SNR). For simplicity, channel gains are assumed
to have the same distribution, and hence it is sufficient to concentrate
on one user to study the performance of interest. In cell $j$, with
available bandwidth $B_{0}$, the wireless downlink rate of a user
is
\begin{equation}
r_{j}=\frac{B_{0}}{U_{j}}\log_{2}\left(1+\frac{P_{t}g_{j}x_{j}^{-\alpha}}{\frac{B_{0}}{U_{j}}\sigma^{2}}\right),\label{eq:r_jk}
\end{equation}
where $P_{t}$ is the BS transmit power, $\sigma^{2}$ denotes the
constant additive noise power, $g_{j}$ is the exponentially distributed
channel power with unit mean, $\alpha$ is the path-loss exponent,
and $x_{j}$ denotes the distance between the target user and the
local BS $j$. To avoid interruptions during the playback of the file
and guarantee the quality of experience (QoE) of users, the downlink
rate $r_{j}$ cannot be lower than the playback rate $r_{0}$. If
$r_{j}\geq r_{0}$, the actual rate dedicated to the user will be
$r_{0}$ (here we assume fixed rate transmission).

A successful file delivery occurs when both the wireless transmission
and the network support meet the requirements. Define the USP of a
user in cell $j$ as $P_{j}^{U}$, which is the probability of the
user successfully obtaining its requested file. Then, we can compute
the success probability by
\begin{equation}
P_{j}^{U}=P_{j}^{W}\cdot P_{j}^{N},\label{eq:P_U}
\end{equation}
where $P_{j}^{W}$ and $P_{j}^{N}$ stand for the success probability
of wireless transmission and network support, respectively. When the
downlink rate reaches the required rate $r_{0}$ (bits/sec/Hz), user
$k$ will enjoy a successful wireless transmission, and thus we have
\begin{equation}
P_{j}^{W}=\mathbb{P}\left[r_{\left(j\right)_{k}}\geq r_{0}\right].\label{eq:P_W}
\end{equation}

As for the network support, if the user's requested file is already
stored in the local cache, the user is considered to be successfully
supported by the network. Otherwise, the requested file needs to be
delivered via the backhaul first. However, the backhaul capacity is
limited and usually not all the users demanding backhaul delivery
can be supported simultaneously. To be specific, in cell $j$, the
largest number of users that can be supported at the same time by
the backhaul is given by $\left\lfloor \left.c_{B,j}\right/r_{0}\right\rfloor $.
After a successful backhaul delivery, the requested file can be transmitted
to the user from BS $j$. Therefore, the network support success probability
is given by
\begin{equation}
P_{j}^{N}=h_{j}+\left(1-h_{j}\right)\cdot P_{j}^{B},\label{eq:P_N}
\end{equation}
where $P_{j}^{B}$ is the probability that the backhaul is available
to a user in cell $j$. We define $B_{j}\triangleq\left\lfloor \left.c_{B,j}\right/r_{0}\right\rfloor $
as the normalized backhaul capacity of BS $j$.

\section{Optimal Cache Size Allocation}

In this section, we will first introduce some auxiliary results, then
give analysis based on a single-cell scenario, and finally investigate
the cache size allocation problem in the multi-cell scenario, for
which the optimal solution will be proposed. In addition, we will
analyze how the optimal allocation scheme is affected by various system
factors, and provide insights for the cache deployment.

\subsection{Auxiliary Results}

In our model, the system performance mainly depends on three factors,
namely, wireless transmission, backhaul delivery and cache hits, which
will critically affect the cache size allocation. We begin with some
auxiliary results for these factors, which will be frequently used
in the remainder of the paper.

\subsubsection{\textmd{Success Probability of Wireless Transmission}}

We assume users are uniformly distributed in the cell, and therefore
the average success probability is given by the following lemma.
\begin{lem}
In the downlink cellular network model, the success probability of
wireless transmission for a user in cell $j$ is
\begin{equation}
P_{j}^{W}=\frac{2}{R_{j}^{2}}\int_{0}^{R_{j}}\exp\left(-\frac{B_{0}\sigma^{2}x^{\alpha}\left(2^{\frac{r_{0}U_{j}}{B_{0}}}-1\right)}{P_{t}U_{j}}\right)x\mathrm{d}x.\label{eq:P_W_lemma}
\end{equation}
\end{lem}
\begin{IEEEproof}
According to the assumption, the probability density function $f_{X_{j}}(x)$
of the distance $x_{j}$ between a typical user and its serving BS
$j$ is $f_{X_{j}}(x)=\frac{2x}{R_{j}^{2}}.$ From  (\ref{eq:r_jk}),
the outage probability of a user in cell $j$ can be given as
\begin{alignat}{1}
 & \mathbb{P}\left[r_{j}<r_{0}\right]=\int_{0}^{R_{j}}\mathbb{P}\left[\left.r_{j}<r_{0}\right|x_{j}=x\right]f_{X_{j}}\left(x\right)\textrm{d}x\nonumber \\
 & =-\frac{2}{R_{j}^{2}}\int_{0}^{R_{j}}\exp\left(-\frac{B_{0}\sigma^{2}x^{\alpha}\left(2^{\frac{r_{0}U_{j}}{B_{0}}}-1\right)}{P_{t}U_{j}}\right)x\textrm{d}x+1.\label{eq:-1}
\end{alignat}
Since $P_{j}^{W}=1-\mathbb{P}\left[r_{j}<r_{0}\right]$, we can obtain
(\ref{eq:P_W_lemma}).
\end{IEEEproof}

\subsubsection{\textmd{Success Probability of Backhaul Delivery}}

Denote the number of users who are requiring backhaul delivery in
cell $j$ as $U_{j}$. For backhaul delivery, it is assumed that if
$U_{j}\leq B_{j}$, all the users can be successfully supported, while
if $U_{j}>B_{j}$, $B_{j}$ users will be randomly picked and get
successful delivery. So the overall success probability of backhaul
delivery is given by the following lemma.
\begin{lem}
If all the users within a cell have an equal chance to be served by
the backhaul, the support probability of backhaul delivery for a user
in cell $j$ is
\begin{equation}
P_{j}^{B}=\sum_{m=0}^{U_{j}-1}\binom{\ensuremath{U_{j}-1}}{m}h_{j}^{U_{j}-1-m}\left(1-h_{j}\right)^{m}\cdot\min\left\{ 1,\frac{B_{j}}{m+1}\right\} .\label{eq:P_B_lemma}
\end{equation}
\end{lem}
\begin{IEEEproof}
Assume that in addition to the target user, there are another $m$
users requiring backhaul service, $m\in\left\{ 0,1,\dots,U_{j}-1\right\} $.
When $m\leq B_{j}-1$, the target user can always be scheduled to
use the backhaul. When $m\geq B_{j}$, the user will get the chance
to use backhaul at a probability of $\frac{B_{j}}{m+1}$. Therefore,
the support probability of backhaul delivery for a user in cell $j$
can be written as
\begin{align}
 & \hspace{1.5em}P_{j}^{B}=\sum_{m=0}^{B_{j}-1}\binom{\ensuremath{U_{j}-1}}{m}h_{j}^{U_{j}-1-m}\left(1-h_{j}\right)^{m}\nonumber \\
 & \hspace{1.5em}\hspace{2.5em}+\sum_{m=B_{j}}^{U_{j}-1}\binom{\ensuremath{U_{j}-1}}{m}h_{j}^{U_{j}-1-m}\left(1-h_{j}\right)^{m}\cdot\frac{B_{j}}{m+1}.\label{eq:p_B_derive}
\end{align}
As a result, (\ref{eq:P_B_lemma}) can be obtained.
\end{IEEEproof}

\subsubsection{\textmd{Cache Hit Ratio}}

In this paper, we adopt Zipf distribution, an empirical model widely
used in related works \cite{CacheSmallCell,cache_coding_Placement},
to model the content popularity. The Zipf distribution states that
the popularity $p\left(f\right)$ of file $f$ is inversely proportional
to its rank, which is written as
\begin{equation}
p\left(f\right)=\frac{\frac{1}{f^{\gamma_{p}}}}{\sum_{g=1}^{F}\frac{1}{g^{\gamma_{p}}}},\label{eq:zipf}
\end{equation}
where $\gamma_{p}$ is the Zipf exponent characterizing the distribution.
Then the cache hit ratio of cell $j$ is given by
\begin{equation}
h_{j}=\frac{\sum_{f=1}^{s_{j}}\frac{1}{f^{\gamma_{p}}}}{\sum_{g=1}^{F}\frac{1}{g^{\gamma_{p}}}}.\label{eq:hit ratio zipf}
\end{equation}

\subsubsection{\textmd{User Success Probability}}

Substituting  (\ref{eq:P_N}), (\ref{eq:P_W_lemma}), (\ref{eq:P_B_lemma})
and (\ref{eq:hit ratio zipf}) into (\ref{eq:P_U}), we can obtain
the expression of USP as
\begin{equation}
P_{j}^{U}=P_{j}^{W}\cdot\left(\frac{\sum_{f=1}^{s_{j}}\frac{1}{f^{\gamma_{p}}}}{\sum_{f=1}^{F}\frac{1}{g^{\gamma_{p}}}}+\left(1-\frac{\sum_{f=1}^{s_{j}}\frac{1}{f^{\gamma_{p}}}}{\sum_{f=1}^{F}\frac{1}{g^{\gamma_{p}}}}\right)\cdot P_{j}^{B}\right).\label{eq:P_U_full}
\end{equation}

\subsection{\label{sub:The-Single-Cell-Scenario:}The Single-Cell Scenario}

We first focus on a single-cell scenario to simplify the analysis.
The subscript $j$ which is used to distinguish cells in the aforementioned
formulas can be ignored in this case. When the backhaul is fixed,
we are interested in the cache size that is required to be allocated
to the local BS in order to reach a USP threshold $\theta_{th}$.
Thus the problem is formulated as minimizing the cache size allocated
to the local BS:
\begin{align*}
\mathcal{Q}\left(\theta_{th}\right):\textrm{minimize} & \hspace{1em}s\\
\textrm{subject to} & \hspace{1em}P^{U}\geq\theta_{th},\forall k\in\left\{ 1,\dots,U\right\} \\
 & \hspace{1em}s\in\mathbb{N}.
\end{align*}
It can be easily checked that $P^{U}$ is monotonically increasing
w.r.t. $s$, so we can adopt a bisection search algorithm to solve
problem $\mathcal{Q}$ with a computational complexity of $\mathcal{O}\left(\log_{2}\left(F\right)\right)$.
In order to gain some insights into the influence of different parameters
on the required cache size, we will further provide a closed-form
expression for $P^{U}$ and an approximation solution for problem
$\mathcal{Q}$.

To begin with, we use the integral form to approximate the popularity
of file $f$ as
\begin{equation}
p\left(f\right)\approx\left.\frac{1}{f^{\gamma_{p}}}\right/\int_{0}^{F}\frac{1}{g^{\gamma_{p}}}\textrm{d}g.\label{eq:popularity_intgrl}
\end{equation}
Then the cache hit ratio is also rewritten in the form of an integral:
\begin{equation}
h=\int_{0}^{s}p\left(f\right)\textrm{d}f.\label{eq:hit ratio_intgrl}
\end{equation}
For $\gamma_{p}\neq1$, we can get
\begin{equation}
h\approx\frac{s^{1-\gamma_{p}}-1}{F^{1-\gamma_{p}}-1}.\label{eq:hit ratio_apprx}
\end{equation}
The result of  (\ref{eq:P_B_lemma}) is relaxed as
\begin{align}
P^{N} & \leq h+\left(1-h\right)\sum_{m=0}^{U-1}\binom{\ensuremath{U-1}}{m}h^{U-1-m}\left(1-h\right)^{m}\frac{B}{m+1}\nonumber \\
 & =h+\frac{B}{U}\left(1-h^{U}\right).\label{eq:P_N_relax}
\end{align}
Since a probability cannot be greater than 1, we have
\begin{equation}
P^{N}\leq\min\left(h+\frac{B}{U}\left(1-h^{U}\right),1\right).
\end{equation}
Note that when $h+\frac{B}{U}\left(1-h^{U}\right)\geq1$, $P^{N}$
may become close to 1, which means that the network (including caching
and backhaul) is always competent to support users, and the USP only
depends on the probability of wireless transmission success, i.e.,
$P^{U}=P^{W}$. However, actually both caching capacity and backhaul
capacity are far from enough for user requirements, i.e., $h+\frac{B}{U}\left(1-h^{U}\right)<1$.
In fact, we usually have $0\leq h<1$ and $h^{U}\ll1$, and thus,
an approximation for $P^{U}$ can be obtained as
\begin{align}
P^{U} & =P_{k}^{W}\cdot P^{N}\leq P^{W}\left(h+\frac{B}{U}\left(1-h^{U}\right)\right)\leq P^{W}\left(h+\frac{B}{U}\right),\nonumber \\
 & \approx P^{W}\left(\frac{s^{1-\gamma_{p}}-1}{F^{1-\gamma_{p}}-1}+\frac{B}{U}\right),\textrm{ for }\mbox{\ensuremath{\gamma_{p}\neq}1}\label{eq:P_U_relax}
\end{align}
which is a closed-form expression for $P^{U}$. From formula (\ref{eq:hit ratio_apprx})
and (\ref{eq:P_U_relax}), a tradeoff between the cache size $s$
and backhaul capacity $B$ can be observed in order to achieve the
required value of $P^{U}$. Given the USP threshold $\theta_{th}$,
we can get the approximated minimum required cache size as
\begin{equation}
s_{\min}\approx\left[\left(\frac{\theta_{th}}{P^{W}}-\frac{B}{U}\right)\left(F^{1-\gamma_{p}}-1\right)+1\right]^{\frac{1}{1-\gamma_{p}}},\textrm{ for }\mbox{\ensuremath{\gamma_{p}\neq}1},\label{eq:S_min}
\end{equation}
from which some key insights can be gained as follows:

1) When $\gamma_{p}>1$ and $F\gg1$, we have $s_{\min}\approx\left(1-\frac{\theta_{th}}{P^{W}}+\frac{B}{U}\right)^{\frac{1}{1-\gamma_{p}}}$.
In this case, $s_{\min}$ increases exponentially with $\theta_{th}$
and $U$, and decreases exponentially with $B$ and $P^{W}$. Note
that $P^{W}$ increases when the cellular wireless channel becomes
favorable and $U$ gets smaller. So a larger $U$ will always result
in a larger $s_{\textrm{min}}$. Additionally, in this scenario, $s_{\min}$
has nothing to do with $F$, which is due to the highly concentrated
popularity distribution, and caching a small subset of files will
be sufficient.

2) When $\gamma_{p}<1$ and $F\gg1$, $s_{\min}\approx\left(\frac{\theta_{th}}{P^{W}}-\frac{B}{U}\right)^{\frac{1}{1-\gamma_{p}}}F$.
In this case, $s_{\min}$ also increases exponentially with $\theta_{th}$
and $U$, and decreases exponentially with $B$ and $P^{W}$. Furthermore,
$s_{\min}$ increases linearly with $F$. This is because the caches
are assumed to store the most popular files, which is the optimal
caching strategy when each user is allowed to access only one BS.
Consequently, when files are of a relatively uniform popularity, in
order to maintain the cache hit ratio, the cache size has to grow
linearly with the file library size.

\subsection{The Multi-cell Scenario}

In the multi-cell scenario, the main objective is to find the optimal
allocation under a given cache budget. For simplicity, each cell is
assumed to operate on an independent bandwidth, and thus the inter-cell
inference is ignored. To achieve fairness among different cells, the
problem is formulated as maximizing the minimum USP with a normalized
cache budget $C_{0}$, and we can obtain the equivalent epigraph form
of the optimization problem as
\begin{align*}
\mathcal{P}':\underset{\left\{ s_{j}\right\} }{\textrm{maximize}} & \hspace{1em}\rho\\
\textrm{subject to} & \hspace{1em}P_{j}^{U}\geq\rho,\forall j\in\left\{ 1,\dots,N_{b}\right\} \\
 & \hspace{1em}\sum_{j=1}^{N_{b}}s_{j}\leq C_{0}\\
 & \hspace{1em}0\leq\rho\leq1\\
 & \hspace{1em}s_{j}\in\mathbb{N},\forall j\in\left\{ 1,\dots,N_{b}\right\} .
\end{align*}
For a fixed $\rho$, we first decouple problem $\mathcal{P}'$ into
subproblems w.r.t. each BS, i.e., a series of problems $\left\{ \mathcal{Q}\left(\rho\right)\right\} $
to find the minimum cache size for each BS, which follows the solution
for the single-cell case. Then we decide whether the total cache size
needed is within our cache budget. Based on this idea, the problem
can be solved by a bisection search procedure, as presented in Algorithm
1.
\begin{algorithm}
\caption{Bisection Search Algorithm for Cache Size Allocation in the Multi-cell
Scenario}

\textbf{Step 0: }Initialize $\rho_{\textrm{low}}=0$, $\rho_{\textrm{up}}=1$,
$\rho_{\textrm{mid}}=0$;

\textbf{Step 1: }Repeat
\begin{enumerate}
\item Set $\rho_{\textrm{mid}}\leftarrow\frac{\rho_{\textrm{low}}+\rho_{\textrm{up}}}{2}$;
\item Solve $N_{b}$ subproblems $\left\{ \mathcal{Q}\left(\rho\right)\right\} $
and obtain the minimum cache size $\left\{ s_{j}\right\} $ for each
subproblem. If $\sum_{j=1}^{N_{b}}s_{j}\leq C_{0}$, set $\rho_{\textrm{low}}=\rho_{\textrm{mid}}$;
otherwise, set $\rho_{\textrm{up}}=\rho_{\textrm{mid}}$;
\end{enumerate}
\textbf{Step 2: }Until $\rho_{\textrm{low}}-\rho_{\textrm{up}}<\epsilon$,
obtain $\rho^{*}=\rho_{\textrm{mid}}$ and obtain the optimal cache
size allocation scheme $\left\{ s_{j}^{*}\right\} _{j=1}^{Nb}$;

\textbf{End }
\end{algorithm}

Our formulation aims at achieving a max-min fairness by the cache
size allocation. There are a lot of factors influencing the allocation.
Since the cellular model considered in this work is backhaul-limited,
we can set all the parameters except the backhaul capacities to be
identical for every cell. From Algorithm 1, as well as the analysis
of the single-cell scenario, some insights for cache deployment have
been determined and are given as follows:

1) When the cache budget is small, the BSs with a lower-capacity backhaul
will obtain a larger quota of the cache size in oder to enhance their
USP to catch up with the other BSs, and some BSs with high-capacity
backhaul links may not be able to get caches.

2) When the cache budget exceeds a certain level, the system performance
will be saturated. This is because the cache hit ratio for every cell
becomes 1 and so does the network support probability, which means
that the cellular network is no longer backhaul-limited.

3) Given a USP threshold, with a smaller $\gamma_{p}$, the cache
budget will be higher, and the influence of the total number of files
will be more significant, and vice versa.

These observations will be verified via simulations in the next section.

\section{Simulation Results}

In this section, we will present simulation results to verify the
effect of cache size allocation and illustrate the impact of various
system parameters. In simulations, we use the cellular network parameters
as shown in Table \ref{tab:Simulation-Parameters}.
\begin{table}
\caption{\label{tab:Simulation-Parameters}Simulation Parameters}

\centering{}%
\begin{tabular}{|c|c|}
\hline
\textbf{Parameter} & \textbf{Value}\tabularnewline
\hline
Cell radius $R_{j}$ & 20 m\tabularnewline
Path-loss exponent $\alpha$ & 4\tabularnewline
Noise power $\sigma^{2}$  & -102 dBm\tabularnewline
Transmit bandwidth $B_{0}$ & 10 MHz \tabularnewline
BS transmit power $P_{t}$ & 1 w\tabularnewline
Data rate requirement $r_{0}$ & 2 Mbps\tabularnewline
\hline
\end{tabular}
\end{table}

\subsection{Backhaul-Cache Tradeoff}

\begin{figure}
\begin{centering}
\includegraphics[width=0.45\textwidth]{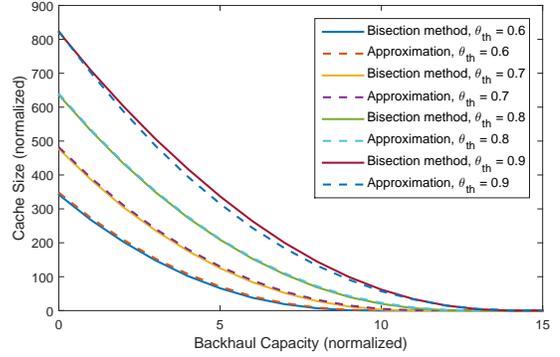}\caption{\label{fig:The-backhaul-caching-tradeoff.}Cache size versus backhaul
capacity.}

\par\end{centering}

\end{figure}

Consider a single cellular network where 15 users make random requests
within a file library of 1000 files following the Zipf distribution
with $\gamma_{p}=0.56$ (suggested by Gill \emph{et al.} \cite{gill2007youtube}).
Fig. \ref{fig:The-backhaul-caching-tradeoff.} explicitly demonstrates
the tradeoff between backhaul capacity and cache size for given success
probability requirements. It can be observed that our approximation
is close to the optimal result, which shows the accuracy of our approximation
in the evaluation. We observe that a higher success probability
can be achieved by increasing either the backhaul capacity or the
cache size. Also, it is shown that when the backhaul capacity is lower,
a backhaul augmentation can reduce the cache size more effectively.

\subsection{Impact of File Popularity Distribution}

\begin{figure}
\begin{centering}
\subfloat[\label{fig:Cache-size-versus F}Cache size versus the number of files.]{\centering{}\includegraphics[width=0.45\textwidth]{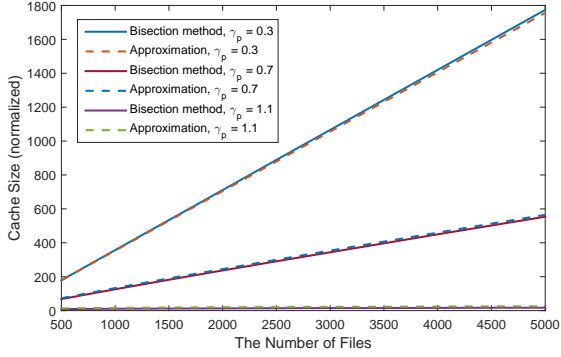}}
\par\end{centering}

\begin{centering}
\subfloat[\label{fig:Cache-size-versus Gamma}Cache size versus popularity shape
parameter.]{\centering{}\includegraphics[width=0.45\textwidth]{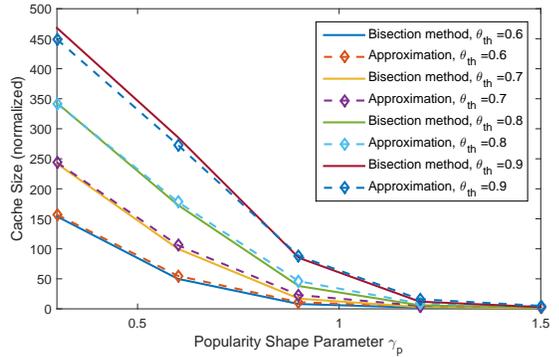}
}
\par\end{centering}

\centering{}\caption{\label{fig:The-influence}The influence of file popularity distribution
on cache size.}
\end{figure}

The file popularity distribution plays a critical role in our system
model. We adopt the Zipf distribution, which depends on two parameters,
the number of files $F$ (i.e., file library size) and the popularity
shape parameter $\gamma_{p}$ (i.e., the Zipf exponent). The effect
of these two parameters on the required cache size is plotted in Fig.
\ref{fig:The-influence}. In Fig. \ref{fig:The-influence}(a), we
set $\theta_{th}=0.8$, and it can be observed that the cache size
increases linearly w.r.t. $F$ when $\gamma_{p}<1$, while the cache
size is close to a constant when $\gamma_{p}>1$. The findings are
in agreement with the insights indicated in Subsection \ref{sub:The-Single-Cell-Scenario:}.

As for the Zipf exponent $\gamma_{p}$, it was estimated as about
0.56 by Gill \emph{et al.} \cite{gill2007youtube} in a campus setting
over three months, and was suggested to be of a higher value of around
1.0 by Cha \emph{et al.} \cite{Cha07Youtube} based on a six-day global
trace of a certain category on YouTube. We investigate the effect
of $\gamma_{p}$ from 0 to $1.5$. A larger $\gamma_{p}$ stands for
a more concentrated popularity distribution. We observe that the required
cache size drops dramatically when $\gamma_{p}$ increases, which
is because a small cache size is enough to support the system, since
a small number of files are responsible for a large number of user
requests.

\subsection{Cache Size Allocation for Different Cache Budget}

In this subsection, we will demonstrate the cache size allocation
scheme under a specific multi-cell scenario consisting of six cells
with backhaul capacities of 0, 2, 6, 10, 20 and 28 Mbps, respectively.
Except the backhaul conditions, the parameters for each cell are set
to be the same as the single-cell case.

\begin{figure}
\centering{}\includegraphics[width=0.45\textwidth]{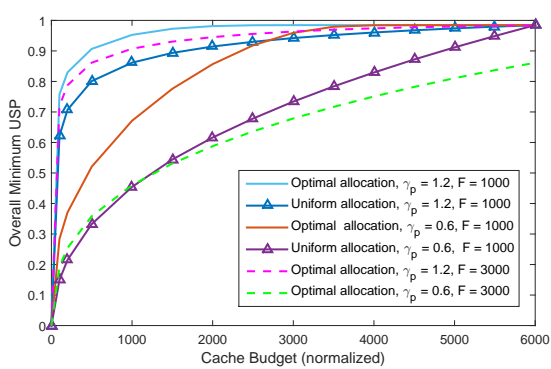}\caption{\label{fig:Comparison}Overall minimum user success probability versus
cache budget. }
\end{figure}

In Fig. \ref{fig:Comparison}, we plot the performance of the optimal
solution and compare it with the uniform allocation of the cache budget.
We evaluate the performance for two different kinds of popularity
distribution: $\gamma_{p}=1.2$, reflecting a more concentrated popularity
distribution, and $\gamma_{p}=0.6$, reflecting a less concentrated
one. It is illustrated that with optimization of the cache budget
allocation, the performance can be significantly improved, which highlights
the importance of cache size allocation. Also, it is noted that the
optimal allocation brings more benefits for a smaller $\gamma_{p}.$
To understand this phenomenon, we can image an extreme popularity
distribution where all users request the same file. In that case,
the uniform allocation will coincide with the optimal allocation.
In addition, it is shown that with a larger $\gamma_{p}$, a lower
cache budget is needed to reach a given USP, and the change in the
file library size will be much more insignificant.

\begin{figure}
\centering{}\includegraphics[width=0.45\textwidth]{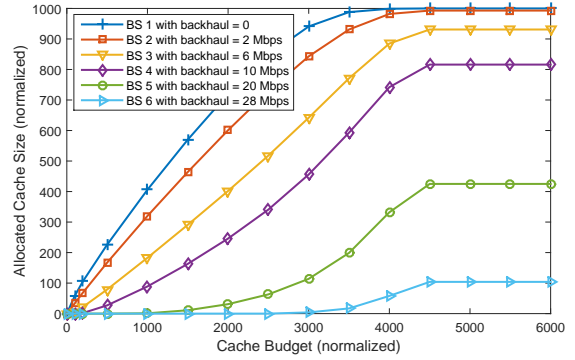}\caption{\label{fig:Allocation}Cache size allocation scheme under different
cache budgets. }
\end{figure}

Fig. \ref{fig:Allocation} shows the details of cache size allocation
under different cache budgets with $\gamma_{p}=0.6$ and $F=1000$.
When the cache budget is small, the BSs with lower backhaul capacities
will have higher priorities to be allocated caches. When the cache
budget is large enough, the performance is saturated, and the allocation
scheme remains the same since a greater cache budget is no longer
useful to improve the system. These behaviors shown in the figure
confirm our initial intuition.

\section{Conclusions}

In this paper, we investigated the cache size allocation problem in
backhaul limited cellular network, while considering the wireless
transmissions and file popularities. In the single-cell scenario,
a closed-form expression of the required cache size to achieve a USP
threshold was derived to illustrate the impacts of different system
parameters. Moreover, the optimum cache size allocation to maximize
the overall system performance was studied for the multi-cell scenario.
It was shown that the optimal allocation can fully exploit the benefits
of caching, and some insights about caching deployment were also discussed.

\bibliographystyle{IEEEtran}
\bibliography{ICC16}

\end{document}